\documentclass[11pt]{article}
	

\usepackage[utf8]{inputenc} 
\usepackage[T1]{fontenc}    
\usepackage{hyperref}       
\usepackage{url}            
\usepackage{booktabs}       
\usepackage{amsfonts}       
\usepackage{amsmath,calc,bm}
\usepackage{nicefrac}       
\usepackage{microtype}      
\usepackage{lipsum}		
\usepackage{graphicx}
\usepackage{natbib}
\usepackage{doi}
\usepackage{pstricks}

\title{Quantum Vacuum Energy of Self-Similar Configurations}

\date{} 					

\author{ \href{https://orcid.org/0000-0002-5575-6775}{\includegraphics[scale=0.06]{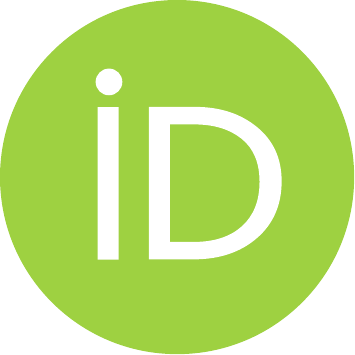}\hspace{1mm}In\'es Cavero-Pel\'aez}\thanks{cavero@unizar.es} 
        \\
		Centro Universitario de la Defensa (CUD), 50090 Zaragoza  \\
		Departamento de F\'isica Te\'orica\\ Facultad de Ciencias, Universidad de Zaragoza\\
		50009 Zaragoza, Spain\\
	\href{https://orcid.org/0000-0003-0812-0421}{\includegraphics[scale=0.06]{orcid.pdf}\hspace{1mm}Prachi Parashar}\thanks{Prachi.Parashar@jalc.edu} \\
	John A. Logan College\\
	Carterville, IL 62918, USA\\ 
	\href{https://orcid.org/0000-0002-4549-5655}{\includegraphics[scale=0.06]{orcid.pdf}\hspace{1mm}K.V. Shajesh}\thanks{kvshajesh@gmail.com} \\
	Department of Physics\\
	Southern Illinois University-Carbondale\\
	Carbondale, IL 62901, USA\\
}

\hypersetup{
pdftitle={Quantum Vacuum Energy of Self-Similar Configurations},
pdfauthor={Cavero-Pel\'aez, I.; Parashar, P.; Shajesh, K.V.},
pdfkeywords={Casimir effect, self-similar systems, smooth manifold, fractal space, spectral functions},
}

\begin{document}
\maketitle

\begin{abstract}
We offer in this review a description of the vacuum energy of self-similar systems. We describe two views of setting self-similar structures and point out the main differences. A review of the authors' work on the subject is presented, where they treat the self-similar system as a many-object problem embedded in a regular smooth manifold. Focused on Dirichlet boundary conditions, we report a systematic way of calculating the Casimir energy of self-similar bodies where the knowledge of the quantum vacuum energy of the single building block element is assumed and in fact already known. A fundamental property that allows us to proceed with our method is the dependence of the energy on a geometrical parameter that makes it possible to establish the scaling property of self-similar systems. Several examples are given. We also describe the situation, shown by other authors, where the embedded space is a fractal space itself, having fractal dimension. A fractal space does not hold properties that are rather common in regular spaces like the tangent space. We  refer to other authors who explain how some self-similar configurations "do not have any smooth structures and one cannot define differential operators on them directly". This gives rise to important differences in the behavior of the vacuum. 
\end{abstract}





\href{https://doi.org/10.3390/universe7050128}{https://doi.org/10.3390/universe7050128}




\section{Introduction}

The concept of the quantum vacuum is a subject that has brought much interest in the community, at least since it acquired a strong mathematical formulation in the context of QED as quantum fluctuations of the electromagnetic fields. However, some decades before that, the black body theory pointed at the concept of zero-point energy. It occurs that the vacuum is anything but calm and gives rise to several observable effects   \cite{miloni_book}. One of the most direct observations of the existence of the quantum vacuum is the so-called Casimir effect in connection with the energy of the vacuum under the influence of certain boundary conditions or external conditions \cite{plunien1986,milton_book, bordag_book, bordag_report, klimchitskaya_overview2006,Farina_CEaspects2006}. In few words, its behavior is as follows. The vacuum spontaneously generates all kind of virtual particles whose fields fluctuate, subject to conditions imposed by  the region where the fluctuations occur. For example, if we consider a bounded region, these fluctuations have limitations in the allowed modes compared to the fluctuations of the fields in the free space, giving rise to a pressure on the boundaries of that region. This effect generates a force that can be measured whenever it implies the interaction between bodies \cite{Lamoreaux2004}. As much as theory and experiment have converged into an agreement on the outcome of this effect, there are still fundamental issues related to how this phenomenon occurs that need to be explained. We still do not understand how factors like the geometry, the materials, or the dimension, change the outcome of the vacuum energy. We cannot predict its value unless an explicit calculation for each particular system is done individually. In this sense, new investigations on those lines can be enlightening. Here, we take a modest look at this field in fractal structures, in particular self-similar~systems.

In quantum field theory, the energy of the ground state of a system is called the zero-point energy. The quantum field can be seen as a system of oscillators with all possible frequencies. It is then possible to calculate the zero-point energy as the infinite sum\footnote{{We assume} $\hbar=c=1$.}
\begin{equation}
    E_0=\frac{1}{2}\sum_n\omega_n,\label{zero-point}
\end{equation}
where the subindex $n$ indicates all possible quantum numbers. When the system is in an embedded space, the oscillation modes of the fluctuating fields change accordingly, and as a consequence, the zero-point energy differs from the one calculated in free space. This difference gives rise to the Casimir energy\footnote{Of course this is a formal definition and must not be understood as a mere difference between two quantities that are, as a matter of fact, infinite. To make sense of it, one must use regularization methods to extract the finite part of such difference.}. 

The quantum vacuum is at the center of all fundamental physics, so it is not surprising that the Casimir effect gets the attention of scientist from different branches of the modern physics. From nanotechnology or new materials to cosmology and astrophysics, the Casimir effect has became an important subject of study. This is due to the nature of the effect, because it involves any quantum field interacting with an external condition as a boundary, a potential or an external field. As a consequence, the free vacuum configuration gets modified, and the vacuum energy can be identified with a self-energy or an interaction energy among objects or other boundaries\footnote{{We do not distinguish for now between global energy and energy density.}}. 
In addition, the resulting energy is also dependent on the space where the vacuum configuration is studied. This encompasses the geometry, topology, or dimension of the space under consideration. Both the finite and the divergent parts of the vacuum energy (the latter one resulting from the divergent nature of the sum in Equation (\ref{zero-point}) involving infinite terms) change accordingly, and therefore it is possible to write an expression for the energy that contains  information about ``the stage where the vacuum is performing''. In other words, the solution of the vacuum energy  in this way reflects the geometry associated with the vacuum under consideration. It is not surprising, since the Casimir energy is intimately related to the spectral properties of the geometry. The idea, of course, resembles the famous problem, ``Can one hear the shape of a drum?'' \cite{Kac}.

There are different approaches to calculate the Casimir energy. All of them have to deal with divergences intrinsic to the concept of quantum vacuum. Each of the approaches, moreover,  has its own predisposition towards a regulation method. In a rough classification of the different approaches, we can differentiate two formulations, the mode summation method and the stress tensor method that can be expressed in terms of the propagators. The first method evaluates the sum of the energy eigenvalues of the vacuum field modes as is formally shown in Equation (\ref{zero-point}). Several spectral functions like the heat kernel, zeta function or the cylinder kernel \cite{fulling2007} gives, under some approximations, the correct interpretation of the sum. It corresponds to these spectral functions to reveal the hidden influence that the geometry has over the total Casimir energy.

In this paper, we aim to differentiate between two ways of investigating  a self-similar system. It can be seen as a many body interaction system where the individual bodies are  set in a particular manner just to build up a self-similar structure. This approach is explained in Section \ref{S2}. There, we describe the method developed in \cite{shajeshPRD94} and extended in \cite{shajeshPRD96}, where we calculate the quantum vacuum energy of some self-similar structures. In the former reference, we use the multiple scattering technique to develop a rather simple-looking method that we verify by reproducing our result using a Green's function approach . In this kind of self-similar arrangement, our N-body interaction formulation allows for $N\rightarrow\infty$. As a result, we obtain sums of infinite series of the same nature as those coming from the geometric structure of the fractal. In this approach, it is crucial to know the self-energy of the primitive building block that constitutes the self-similar system.

In Sections \ref{S3} and \ref{S4}, we report new results from applying the method explained in Section \ref{S2}. Section \ref{S3} presents the calculation of the vacuum energy of a configuration of self-similar spheres. Section \ref{S4} reports the results for quasi-periodic configuration plates using plates that satisfy Dirichlet and Neumann boundary conditions.

We give in Section \ref{S5} a review of the construction of some spectral functions
corresponding to the Laplacian operators and boundary conditions that describe, among other things, the building blocks of the self-similar structures we talk about in the previous sections defined in a regular smooth space.

In Section \ref{S6}, on the other hand, we look at the self-similar structure as a fractal space~\cite{Strichartz_book}. This field has not been developed thoroughly in the Casimir community, maybe because it is not so straightforward as to give an interpretation to the convergence ``or not'' of the spectral functions in such a space compared to information these functions give us when they are calculated in the normal regular manifold \cite{Kigami_1989}. We give a small overview of what other authors have investigated about this situation in some particular cases.

Throughout the paper, we concentrate on scalar fields satisfying Dirichlet boundary conditions since that simplifies the calculation we want to illustrate. Only occasionally, we will use Neumann boundary conditions, but in general, one can proceed  in the same manner with either of them.

\section{Self-Similarity as Many-Body Systems on a Regular Smooth Manifold}\label{S2}

The least many-body system one can have is a system of two. The energy of such a system has been broadly studied. In particular, in the last two decades, the application of multiple scattering techniques \cite{Balian_AnnPhys104, Balian_AnnPhys112} to the calculation of the Casimir energy has allowed the study of the vacuum energy of systems made of compact disjoint bodies. It is found out that the total vacuum energy in those cases can split into different terms corresponding to the single-body contributions and the interaction energy between the bodies \cite{K&K_PRL97, Emig_PRD77, Milton_JPhysA41}. Specifically for two disjoint parallel plates at a distance $a$, the total energy per unit area can be expressed as
\begin{equation}
{\cal E} = {\cal E}_0 + \Delta {\cal E}_1 + \Delta {\cal E}_2 + \Delta {\cal E}(a),
\label{CE_decomposition}
\end{equation}
where ${\cal E}_0$ is the energy of the vacuum in the absence of the two objects, $\Delta {\cal E}_i= {\cal E}_i- {\cal E}_0$, $i=1,2$, are the one-body energies associated to the individual objects, and $\Delta {\cal E}(a)$ is the interaction energy per unit area of the plates. In general, the one-body energies and the bulk energy ${\cal E}_0$ diverge, and the Casimir interaction energy per unit area $\Delta {\cal E}(a)$ is finite and is distinctly isolated by its dependence on the distance $a$, a signature of the interaction between the two plates.

A generalization of the two-body interaction of more than two bodies was given in Reference~\cite{Wirzba_1999,Schaden_EPL94} in the context of multiple-scattering techniques and in References~ \cite{Shajesh_PRD83, Shajesh_IntJModPhys14} using Green's function method, but
explicit solutions for the Green functions were reported only for 
configurations with three bodies. In \cite{shajeshPRD94},
the authors find solutions to the Green functions for four bodies, and then they
go further and express the solution to the Green function for $N$ bodies 
as a recursion relation in terms of the Green's functions for $(N-2)$ bodies.
This procedure then lets them extend their solutions for the Green functions
for an infinite sequence of objects by taking the limit $N\to\infty$.

In any of the above cases, two bodies, N bodies, or in the limit $N\to\infty$, they all have in common that the Casimir energy is due to vacuum fluctuations on a regular, smooth manifold regardless of the shape of the interacting bodies. Fixing the shape of the interacting bodies and taking as many of them as are needed, if we now place them in a recurrent manner, we construct a self-similar structure that can be understood as a many-body system. See, for example, any of the figures from {Figure~1} to {Figure 5}.

The property of the self-similarity is well illustrated in the infinite sum of some series. Consider for example
\begin{equation}
x=1+\frac{1}{2} +\frac{1}{4} +\frac{1}{8} + \ldots.
\label{series1}
\end{equation}

Because of the self-similarity concept, formally, the series allows the following identification,
\begin{equation}
x = 1 + \frac{1}{2} x,
\label{summ1}
\end{equation}
which immediately leads to the conclusion that the sum of the series is $x=2$.
We can extend this idea of self-similarity to ``sum'' a divergent series too.
For example, for the divergent sum 
\begin{equation}
x=1+2+4+8+\ldots, 
\label{series2}
\end{equation}
using the idea of self-similarity, we can identify the relation 
\begin{equation}
x=1+2x,
\label{summ2}
\end{equation}
which assigns the value $x=-1$ to the above divergent sum
and is interpreted as the ``sum'' of the divergent series \cite{hardy1956}. At  first glance, it appears awkward to think that a sum of positive real numbers could generate a negative real number. However, in quantum field theory, we encounter regularization procedures like the Riemann zeta function $\zeta(s)$, where this kind of behavior is common and well accepted:
\begin{equation*}
    \zeta(s)=\sum_{n=1}^\infty \frac{1}{n^{-s}},\qquad \zeta(0)=-\frac{1}{2},\qquad \zeta(-1)=-\frac{1}{12}.
\end{equation*}

Next we construct self-similar configurations of several objects, using the idea of self-similarity along the lines of the illustrations above, and derive the Casimir interaction energies for these configurations.
%
\subsection{Self-Similar Parallel Plates}
\begin{figure}[h]
\begin{center}
\includegraphics[width=6.5 cm]{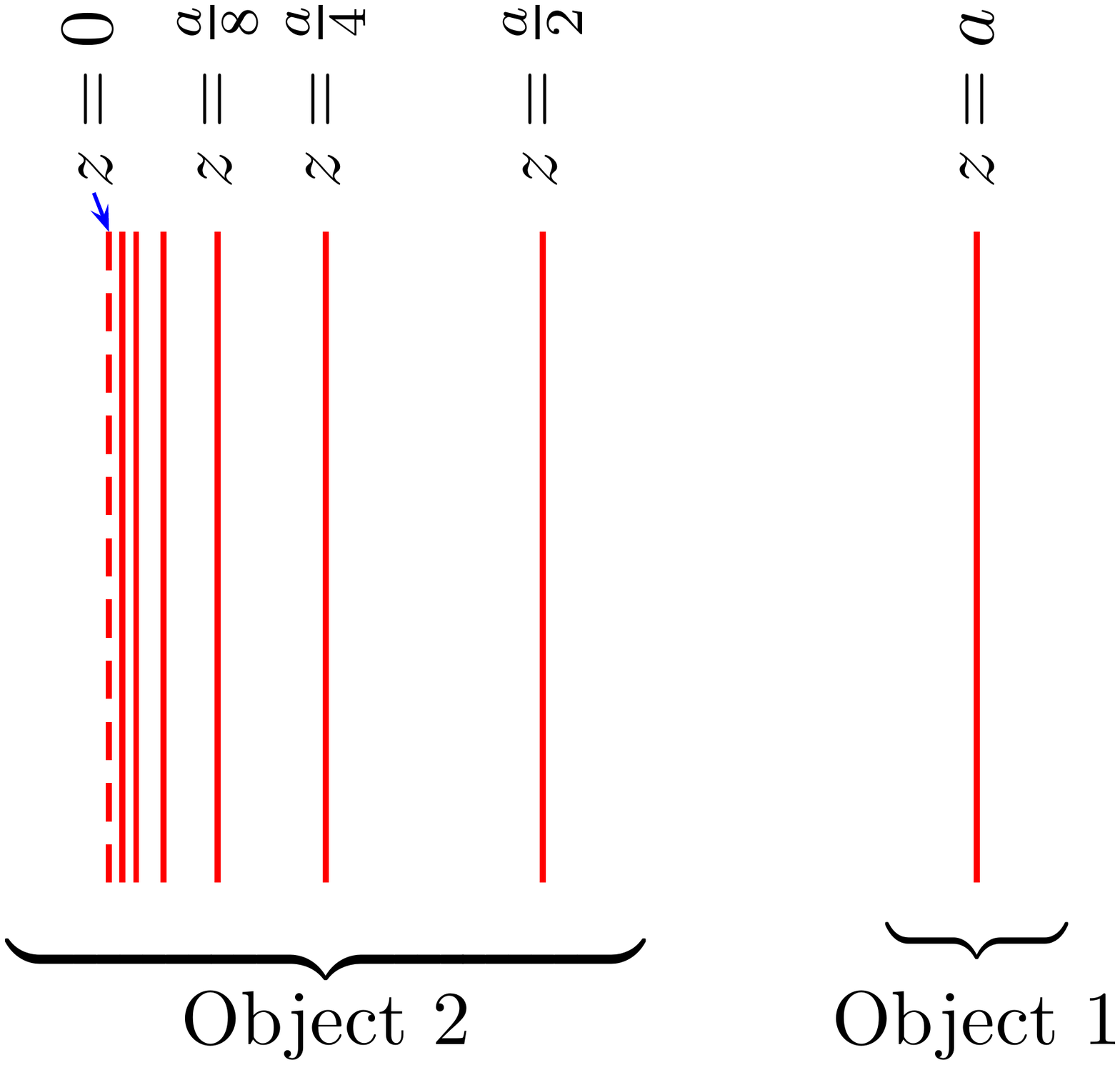}
\caption{A geometric sequence of parallel plates. 
The position of the plates is given by the sequence
$z=a, \frac{a}{2}, \frac{a}{4}, \frac{a}{8}, \ldots$.
The first seven plates of the infinite sequence have been shown.
The dashed line to the left is the limit of this sequence of plates. 
}
\label{fig-a2i-plates-12}
\end{center}
\end{figure}

%
Let us consider an infinite sequence of plates placed at the following positions (see Figure~\ref{fig-a2i-plates-12})
\begin{equation}
z=a, \frac{a}{2}, \frac{a}{4}, \frac{a}{8}, \ldots,
\label{zpos-gpd}
\end{equation}
 such that the distances between the plates successively decrease by a factor of~two.
Let us analyze the energy break up of this infinite sequence of plates by interpreting the single plate at $z=a$ as Object 1 and the rest of the plates to constitute Object 2. One can straight away make the transcription of Equation (\ref{CE_decomposition}) and write
\begin{equation}
    {\cal E} = {\cal E}_0 + \Delta {\cal E}_{O1} + \Delta {\cal E}_{O2} + \Delta {\cal E}_{12}(a).
\label{CE_decomposition_total_stack}
\end{equation}

Apparently, we have reduced the many-body problem to a two-body problem where the interaction term is just $\Delta{\cal E}_{12}(a)$. However the energy of ``the single body'' that we called Object 2, $\Delta{\cal E}_{O2}$, has implicit the interaction energy of the whole pile of plates from $z=0$ to $z=a/2$. We also need to account for that energy, since we are looking for the interaction energy of the whole bunch. We are then interested in expressing the above equation in terms of the individual plates in order to isolate  the whole interaction term that, on the other hand, we know  has to be finite when the interacting bodies are disconnected. Realizing that
\begin{equation}
    \Delta{\cal E}_{O1}= \Delta{\cal E}_1,\qquad \Delta{\cal E}_{O2} = {\cal E}_{O2}-{\cal E}_0,
\end{equation}
we make an analysis of the kind shown in Equations (\ref{series1}) to (\ref{summ2}).
Using the decomposition of energy in Equation (\ref{CE_decomposition}) that for clarity we have re-written in (\ref{CE_decomposition_total_stack}), we find
\begin{eqnarray}
\left( {\cal E}_0+\sum_{i=1}^\infty \Delta {\cal E}_i +\Delta {\cal E}(a)\right) = {\cal E}_0 + \Delta {\cal E}_1+ \left( \sum_{i=2}^\infty \Delta {\cal E}_i + \Delta {\cal E}(a/2) \right) +\Delta {\cal E}_{12}(a), \hspace{5mm}
\label{finCeay}
\end{eqnarray}
where we have isolated the single-body contributions to the energy explicitly. The single-body contributions, in this manner, cancel out in Equation~(\ref{finCeay}) to give
\begin{equation}
\Delta{\cal E}(a) = \Delta{\cal E}(a/2) + \Delta{\cal E}_{12}(a),
\label{gp-ce-aa212}
\end{equation}
which requires some elaboration because we have used the idea of self-similarity in writing Equation~(\ref{gp-ce-aa212}).
The interaction energy of the complete stack of plates in Figure~\ref{fig-a2i-plates-12} is on the left side of Equation~(\ref{gp-ce-aa212}). This is the term of interest for us, and the notation indicates that we consider the interaction energy of a self-similar plate configuration of total width $a$. The first term on the right of Equation~(\ref{gp-ce-aa212}) is the interaction energy of the plates constituting Object 2 in Figure~\ref{fig-a2i-plates-12}\footnote{Note that, according to our notation, the self-similar structure has a width of $a/2$ now.}. The second term on the right of Equation~(\ref{gp-ce-aa212}) is the interaction energy between Object 2 and Object 1. The idea of self-similarity has been used to note that the energy of Object 2 is equal to the energy of the complete stack evaluated for a re-scaled parameter, here $a/2$. 

The interaction energy is a function of the parameter $a$. However, in general, it can also depend on the potentials that describe the plates or the boundary conditions on them,  thereby becoming very hard to determine. Our assumption is that the plates are Dirichlet plates because they satisfy Dirichlet boundary conditions. In that case, the interaction energy is a function of $a$ alone, because that is the only parameter in the problem. On dimensional grounds, we can argue that 
\begin{equation}
\Delta {\cal E}(a/2)=2^3\,\Delta {\cal E}(a).
\label{scaArcs}
\end{equation}

Using the scaling argument of Equation~(\ref{scaArcs}) in Equation~(\ref{gp-ce-aa212}), we identify the relation involving the Casimir interaction energy of the infinite sequence of plates in Figure~\ref{fig-a2i-plates-12},
\begin{equation}
\Delta {\cal E}(a) = 8\,\Delta{\cal E}(a) +\Delta{\cal E}_{12}(a),
\label{1ea=8ea+12ea}
\end{equation}
which is the analog of the relation for infinite series in Equation~(\ref{summ1}) or (\ref{summ2}), here for the Casimir interaction energies.

The relation in Equation~(\ref{1ea=8ea+12ea}) allows us to evaluate $\Delta{\cal E}(a)$ in terms of the interaction energy between Object 1 and Object 2 given by $\Delta{\cal E}_{12}(a)$. As we mentioned, in general, it is a difficult task to evaluate the interaction energy $\Delta{\cal E}_{12}(a)$. But if each of the plates that make up the stack is a Dirichlet plate, the  analysis becomes remarkably simplified because a Dirichlet plate physically disconnects the two spaces across it. In a rigorous way,  explicit decomposition of the total energy in terms of single-body, two-body, and three-body energies, and how they conspire such that the Casimir interaction energy is given completely in terms of interaction of two Dirichlet plates, was described in detail in Reference\,\cite{Shajesh_PRD83}. As a consequence, each Dirichlet plate can only interact with its closest neighbor on the left and on the right. 
We know that the Casimir interaction energy of two Dirichlet plates \cite{milton_book,  Elizalde_AnnJPhys59} separated by distance $a$ is
\begin{equation}
\Delta{\cal E}_{12}(a) =  -\frac{\pi^2}{1440 a^3}.
\label{interaction_En_ParllPlates}
\end{equation}

Thus, the Casimir interaction energy between the two objects separated by distance $a/2$, which is the distance between the plates located at $z=a$ and $z=a/2$ in Figure~\ref{fig-a2i-plates-12}, is obtained by replacing $a$ by $a/2$ in Equation~(\ref{interaction_En_ParllPlates}). The interaction energy of the system is
\begin{equation}
\Delta{\cal E}(a) = 8\,\Delta{\cal E}(a) -\frac{\pi^2}{1440 (a/2)^3},
\end{equation}
which immediately leads to the Casimir interaction energy per unit area for the complete stack in Figure~\ref{fig-a2i-plates-12} given by
\begin{equation}
\Delta{\cal E}(a) = +\frac{8}{7}\frac{\pi^2}{1440 a^3}.
\label{cie-ingp}
\end{equation}

Thus, using the idea of self-similarity, in a self-contained derivation, we have derived the Casimir interaction energy of an infinite stack of plates. Remarkably, the sign of the Casimir interaction energy for this configuration is positive. Thus, the tendency for the infinite sequence of plates in Figure~\ref{fig-a2i-plates-12} is to inflate due to the pressure of vacuum.

\begin{figure}[h]
\begin{center}
\includegraphics[width=9.5 cm]{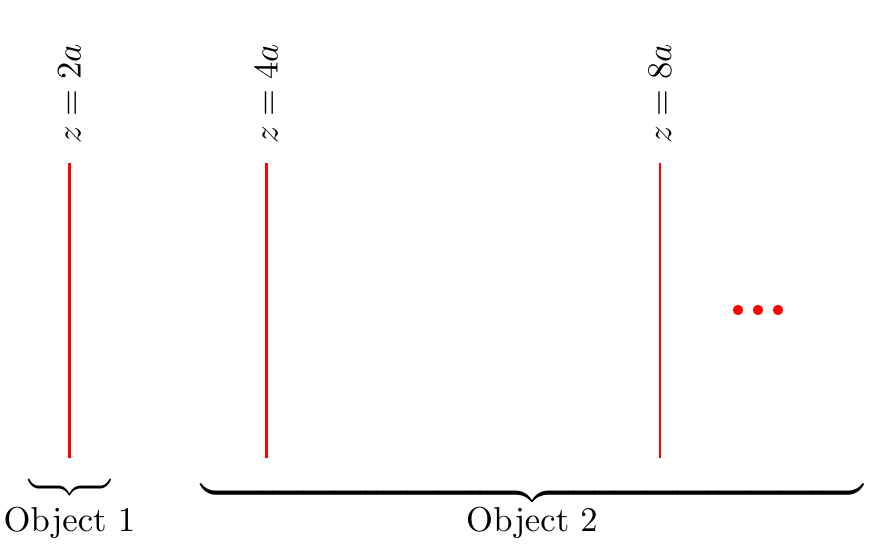}
\caption{A geometric sequence of parallel plates. 
The position of the plates is given by the sequence
$z=2a, 4a, 8a, 16a, \ldots$.}
\label{fig-a2i-plates-in}
\end{center}
\end{figure}

We consider another example
to point out that the Casimir interaction energy
is not always positive for an infinite sequence of plates.
We consider an infinite sequence of plates placed at the following positions
\begin{equation}
z=2a,4a,8a,16a,\dots,
\label{ser-gp-in}
\end{equation}
as described in Figure~\ref{fig-a2i-plates-in}.
We start from $z=2a$ because it extends the series in Equation~(\ref{zpos-gpd})
and later allows us to merge the two stacks, the present and the previous one.
Using the idea of self-similarity, we identify the relation
\begin{equation}
\Delta{\cal E}(2a) = \Delta{\cal E}(4a) -\frac{\pi^2}{1440 (2a)^3},
\end{equation}
where the term on the left is the interaction energy of the whole stack of plates from position $z=2a$, the first term on the right is the interaction energy of the whole stack of plates from position $z=4a$, and $2a$ in the denominator of the second term is the distance between Object 1 and Object 2.
Then, using
\begin{equation}
\Delta{\cal E}(4a) = \frac{1}{2^3}\,\Delta{\cal E}(2a), 
\end{equation}
we immediately learn that
\begin{equation}
\Delta{\cal E}(2a) = -\frac{1}{7}\frac{\pi^2}{1440 a^3}.
\label{cie-degp}
\end{equation}

This suggests that the tendency of the stack of plates 
in Figure~\ref{fig-a2i-plates-in} is to contract under the pressure of vacuum.

\begin{figure}[h]
\begin{center}
\includegraphics[width=12 cm]{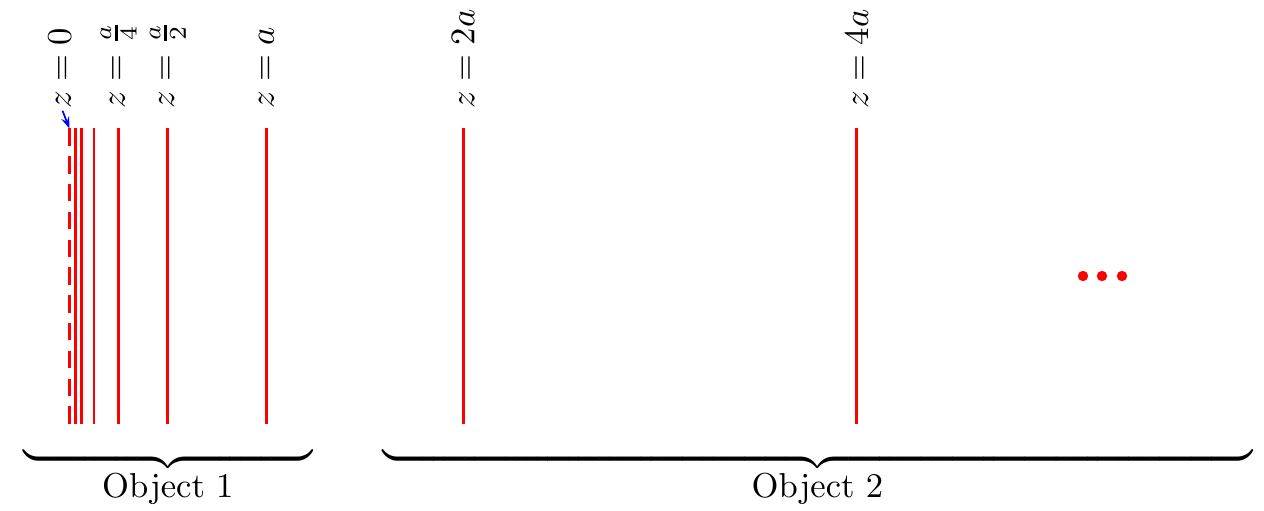}

\caption{A geometric sequence of parallel $\delta$-function plates.
The position of the plates is given by the sequence
$z= \ldots, \frac{a}{8}, \frac{a}{4}, \frac{a}{2}, a, 2a,4a,8a,\dots$.
}
\label{fig-a2i-plates-double}
\end{center}
\end{figure}

Having derived the Casimir interaction energy of two independent stacks,
we now place them such that they can be imagined to form
a sequence that extends on both ends, given by
\begin{equation}
z= \ldots, \frac{a}{8}, \frac{a}{4}, \frac{a}{2}, a, 2a,4a,8a,\dots,
\label{zpos-gpd-do}
\end{equation}
as described in Figure~\ref{fig-a2i-plates-double}.
Since we already derived the energies for the individual stacks,
we can calculate the energy of the complete stack using the two-body
break up of the Casimir energies. Thus, we have the total interaction energy
of the two stacks $\Delta{\cal E}_\text{tot}(a)$ given by the relation
\begin{equation}
\Delta{\cal E}_\text{tot}(a) = \Delta{\cal E}(a) + \Delta{\cal E}(2a)
-\frac{\pi^2}{1440 a^3},
\end{equation}
where the first term on the right is the Casimir interaction energy
$\Delta{\cal E}(a)$ of the first stack given by Equation~(\ref{cie-ingp}),
the second term  is the Casimir interaction energy
$\Delta{\cal E}(2a)$ of the second stack given by Equation~(\ref{cie-degp}),
and the third term is the interaction energy of the two stacks
given by the energy of two Dirichlet plates in Equation~(\ref{interaction_En_ParllPlates}).
Together, we have
\begin{equation}
\Delta{\cal E}_\text{tot}(a) = +\frac{8}{7} \frac{\pi^2}{1440 a^3}
-\frac{1}{7} \frac{\pi^2}{1440 a^3} -\frac{\pi^2}{1440 a^3} =0,
\end{equation}
which suggests that the Casimir energy of the two stacks, in
conjunction, in Figure~\ref{fig-a2i-plates-double},
is exactly zero. Apparently, the pressure due to vacuum that tends
to inflate the first stack when in isolation
and contract the second stack in isolation, when in conjunction,
conspire to balance these opposite tendencies exactly.
It can be easily verified that this cancellation is independent
of the particular choice of breakup into Objects 1 and 2,
which is a signature of self-similarity.

\begin{figure}[h]
\begin{center}
\includegraphics[width=7.5 cm]{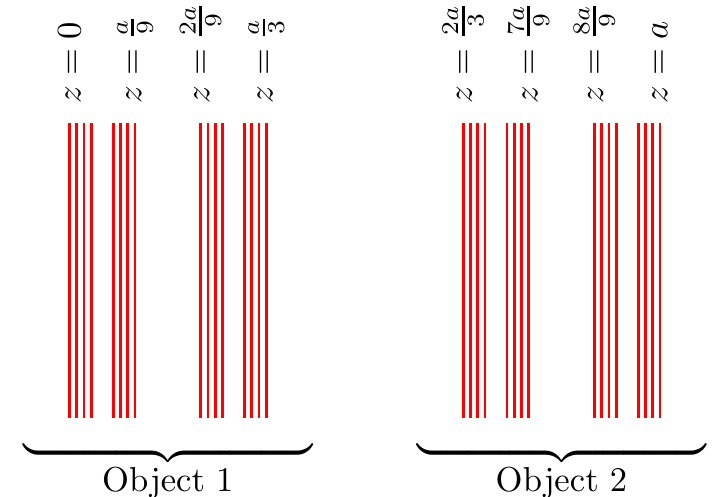}
\caption{A sequence of parallel plates positioned at the points
forming the Cantor set. The figure shows plates positioned at points
generated in four iterations.
}
\label{fig-cantor-plates}
\end{center}
\end{figure}

In our last example with parallel plates, we highlight  
a self-similar configuration of plates motivated from 
the Cantor set. We place a $\delta$-function plate at every point
of the Cantor set. The classic Cantor set is obtained by 
repeatedly  dividing a line segment into three parts and deleting
the central region each time.
We build our stack of plates by placing a $\delta$-function plate
at the edge of the remaining segments in each iteration; see Figure~\ref{fig-cantor-plates}.
The idea of self-similarity and the two-body break-up of energy then 
leads to the relation, in the Dirichlet limit,
\begin{equation}
\Delta{\cal E}(a) = \Delta{\cal E}(a/3) + \Delta{\cal E}(a/3) 
-\frac{\pi^2}{1440 (a/3)^3}.
\end{equation}

Then, using
\begin{equation}
\Delta{\cal E}(a/3) = 3^3\,\Delta{\cal E}(a)
\end{equation}
we have the Casimir interaction for the configuration in 
Figure~\ref{fig-cantor-plates} given by
\begin{equation}
\Delta{\cal E}(a) = +\frac{27}{53}\frac{\pi^2}{1440 a^3}.
\end{equation}

The positive sign signifies that the pressure due to vacuum tends to 
inflate the stack in Figure~\ref{fig-cantor-plates}.

\subsection{Sierpinski Triangles}
%
Using the same scaling arguments, we derive the Casimir energy for a Sierpinski triangle constructed by arranging triangular cavities with perfectly conducting boundaries. Furthermore, we consider these triangular cavities to be the cross section of infinitely long cylinders. We study the Casimir effect of the self-similar structure that emerges from this set up, where the oscillating modes are calculated  in the vacuum space inside the triangular cavity. The arguments given here are developed in Reference~\cite{shajeshPRD96}.

\begin{figure}[h]
\begin{center}
\includegraphics[width=5.5 cm]{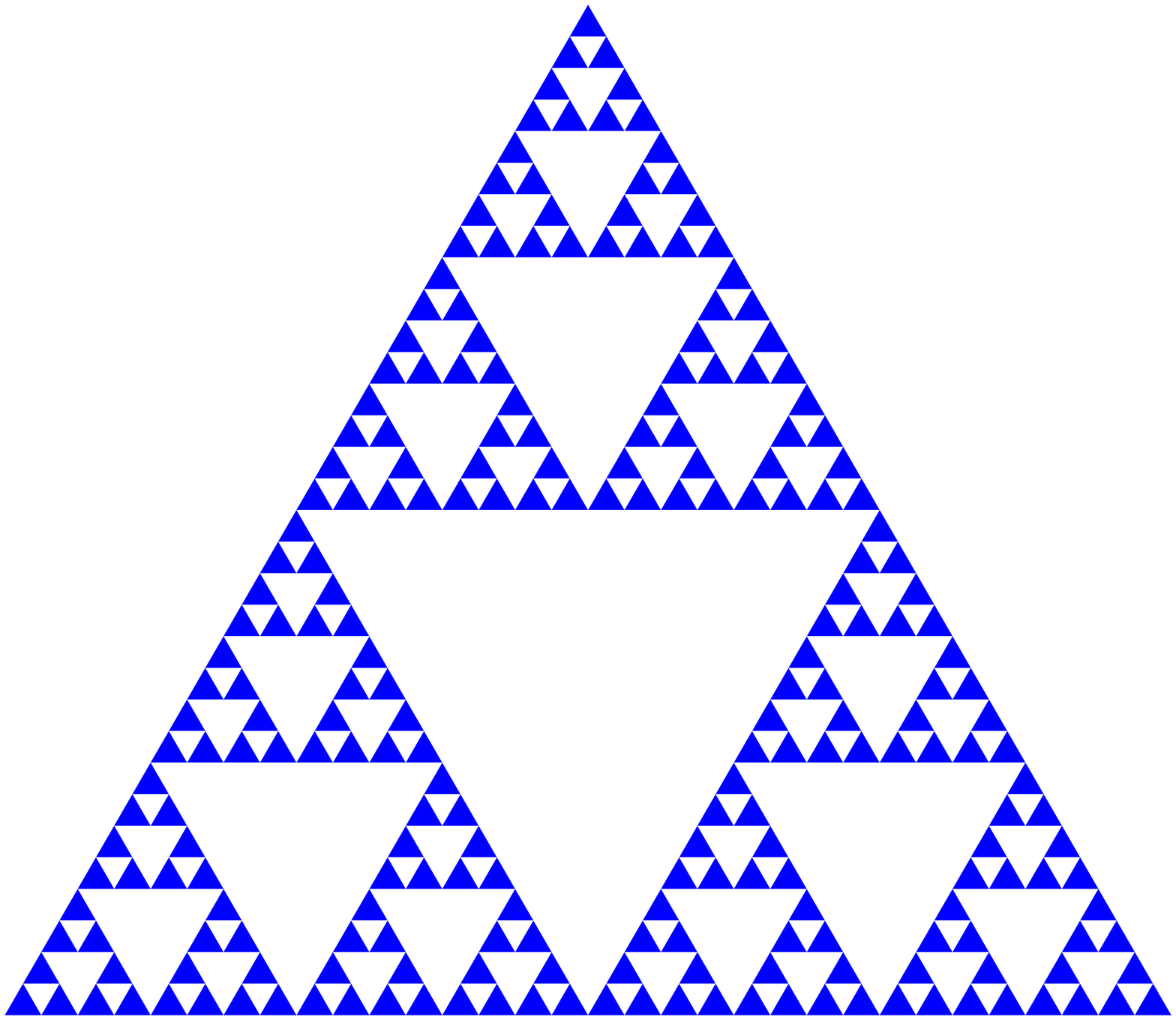}
\caption{Sierpinski triangle. The white regions in the interior are triangular cavities, each of which contributes to the Casimir energy per unit lenght of the Sierpinski triangle. The matter bounding each of the triangles (in blue) is perfectly conducting for the case of electromagnetic fields.}
\label{Sierpinski_triangle}
\end{center}
\end{figure}

Using the decomposition of Casimir energies into single-body energy
and the respective interaction energy between the bodies~\cite{Shajesh_PRD83},
the Casimir energy per unit length of a Sierpinski triangle ${\cal E}_s(a)$
can be decomposed as
\begin{equation}
{\cal E}_s(a) = {\cal E}_\text{int}\left(\frac{a}{2}\right) 
+ 3{\cal E}_s\left(\frac{a}{2}\right),
\label{st-ed3}
\end{equation}
where ${\cal E}_s(a/2)$ is the single-body Casimir energy of each of the three
Sierpinski triangles of side $a/2$ in Figure~\ref{Sierpinski_triangle} and
${\cal E}_\text{int}(a)$ is the interaction energy between the 
three Sierpinski~triangles. 

Arguably, in general, the interaction energy ${\cal E}_\text{int}$
depends on both the interior and exterior modes. However, the Casimir energies
of the cavities are all due to interior modes.
Thus, for consistency, we shall presume the interaction energy 
involves only the interior modes. We shall further justify the 
consistency of this presumption in the following discussion.
Using Equation\,(\ref{st-ed3}) recursively,  we obtain the series
\begin{equation}
{\cal E}_s(a) = {\cal E}_\text{int}\left(\frac{a}{2}\right) 
+ 3\,{\cal E}_\text{int}\left(\frac{a}{2^2}\right) 
+ 3^2\,{\cal E}_\text{int}\left(\frac{a}{2^3}\right) +\ldots.
\label{steint}
\end{equation}

Thus, the evaluation of the Casimir energy reduces to computing the 
interaction energy ${\cal E}_\text{int}$. 

\begin{figure}[h]
\begin{center}
\includegraphics[width=5.5 cm]{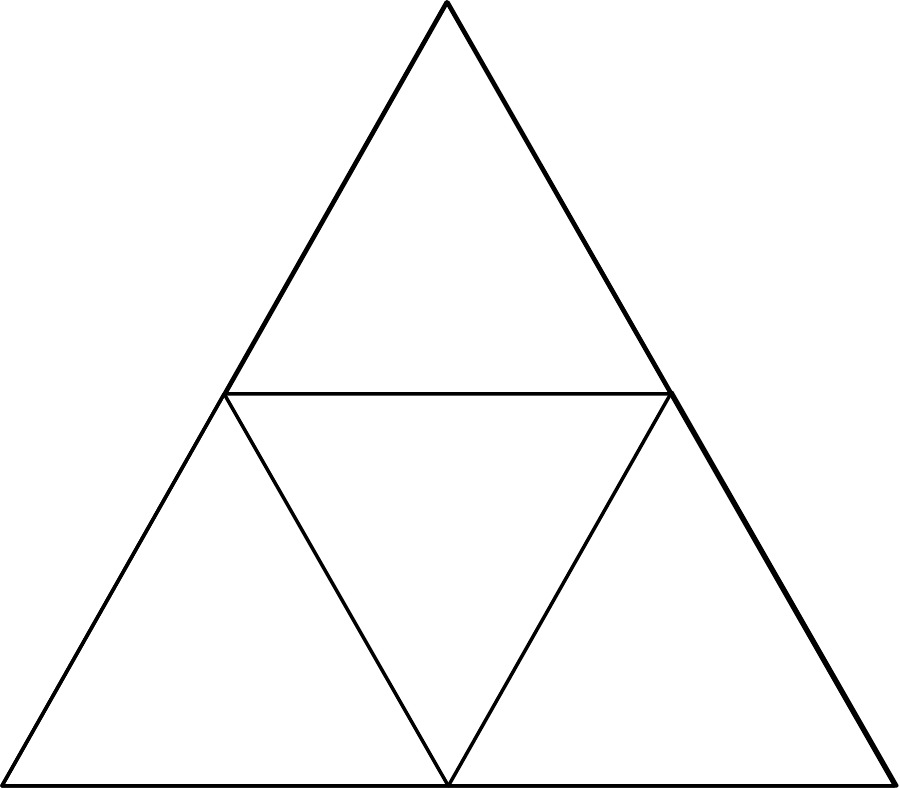}
\caption{Four triangles constituting the Sierpinski triangle.}
\label{fig-fourT}
\end{center}
\end{figure}

Dirichlet boundary conditions require a scalar field to be zero
on the boundary. In our case, this restriction essentially separates the 
physical phenomena on the two sides of the boundary.
Thus, the modes and the associated physical phenomena inside 
Dirichlet cavities are essentially independent of its surroundings. 
Extending this argument to Sierpinski triangles, we learn that 
the interaction energy between two or more Sierpinski triangles is 
independent of the internal structure of each of the Sierpinski triangles.
We can thus infer that the interaction energy of the three 
Dirichlet Sierpinski triangles in Figure~\ref{Sierpinski_triangle}
is identical to the interaction energy the three Dirichlet triangles
in Figure~\ref{fig-fourT}.
We can determine the total energy of the triangles in 
Figure~\ref{fig-fourT} in two independent methods.
In the first method, we argue that the energy is the sum of the four
triangular cavities, $4{\cal E}_\Delta\left(\frac{a}{2}\right)$.
In the second method, we argue that the total energy is the sum 
of the energies of the three outer triangles,
$3{\cal E}_\Delta\left(\frac{a}{2}\right)$,
plus the interaction energy ${\cal E}_\text{int}(a/2)$ 
between the three triangles. That is,
\begin{equation}
4{\cal E}_\Delta\left(\frac{a}{2}\right) 
=3{\cal E}_\Delta\left(\frac{a}{2}\right) 
+ {\cal E}_\text{int}\left(\frac{a}{2}\right). 
\end{equation}

This immediately suggests that the interaction energy of three outer
triangles is completely given by the energy of the inner triangle, 
\begin{equation}
{\cal E}_\text{int}(a) = {\cal E}_\Delta\left(\frac{a}{2}\right) 
= 2^2 {\cal E}_\Delta(a),
\label{est-eint4}
\end{equation}
where in the second equality, we used the
fact that the Casimir energy of a triangle ${\cal E}_\Delta(a)$ scales
like the inverse square of length.
Using Equation\,(\ref{est-eint4}) in Equation\,(\ref{steint}), we derive the Casimir
energy of the Dirichlet Sierpinski triangle in terms of the Casimir energy
of the equilateral triangle as 
\begin{equation}
{\cal E}_s(a) = -\frac{4}{11} {\cal E}_\Delta(a)
\end{equation}
using the divergent sum
\begin{equation}
1 +12 +12^2 +\ldots = -\frac{1}{11}.
\end{equation}
\vspace{-12pt}

\section{Concentric Spheres \textendash  Self-Similar Spheres}\label{S3}

\begin{figure}[h]
\begin{center}
\includegraphics[width=5 cm]{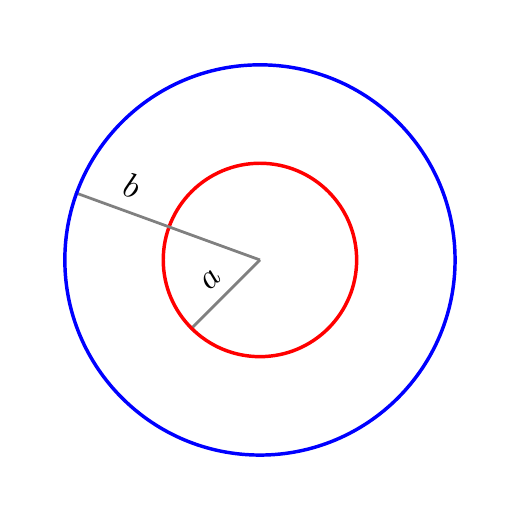}

\caption{Perfectly conducting concentric $\delta$-function spheres.}
\label{fig-con-dpheres}
\end{center}
\end{figure}

We shall now consider a self-similar system built by concentric $\delta$-function spheres.
Using Equations\,(5.6) and (5.7) of Reference\,\cite{Parashar_PRD96},
the Casimir interaction energy between two perfectly conducting
concentric $\delta$-function spheres pictured in Figure\,\ref{fig-con-dpheres}
can be expressed in the form
\begin{equation}
E(a,b) = \frac{1}{a} A\left( \frac{b}{a} \right),
\end{equation}
in terms of
\begin{equation}
A(x) = \frac{1}{2} \int_{-\infty}^\infty \frac{ds}{2\pi}
\sum_{l=1}^\infty (2l+1)
\ln \left[1 - \frac{i_l(s)}{k_l(s)} \frac{k_l(sx)}{i_l(sx)} \right]
\left[1- \frac{\bar i_l(s)}{\bar k_l(s)}
\frac{\bar k_l(sx)}{\bar i_l(sx)} \right].
\hspace{5mm}
\end{equation}

Here $\text{i}_l(t)$ and $\text{k}_l(t)$ are modified spherical Bessel
functions that are related to the modified Bessel functions by the {relations}
\begin{equation}
\text{i}_l(t) = \sqrt{\frac{\pi}{2t}} I_{l+\frac{1}{2}}(t), 
\end{equation}
\begin{equation}
\text{k}_l(t) = \sqrt{\frac{\pi}{2t}} K_{l+\frac{1}{2}}(t).
\end{equation}

In particular $\text{i}_l(t)$,
the modified spherical Bessel function of the first kind,
together with $\text{k}_l(t)$, form a suitable
pair of solutions in the right half of the complex plane
\cite{NIST:DLMF}.
We used bars to define the following operations on the
modified spherical Bessel functions
\begin{equation}
\bar{\text{i}}_l(t)
= \bigg( \frac{1}{t} + \frac{\partial}{\partial t} \bigg) \text{i}_l(t), 
\end{equation}
\begin{equation}
\bar{\text{k}}_l(t)
= \bigg( \frac{1}{t} + \frac{\partial}{\partial t} \bigg) \text{k}_l(t).
\label{bar-mBf}%
\end{equation}%
\unskip
\begin{figure}[h]
\begin{center}
\includegraphics[width=5 cm]{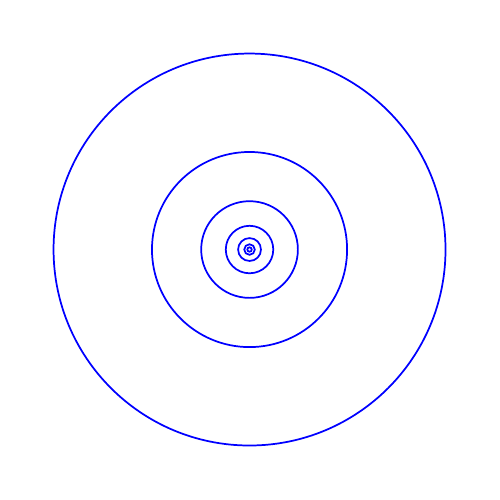}
\caption{{Perfectly} conducting concentric self-similar spheres.}
\label{fig-selfsim-in}
\end{center}
\label{fig-con-n-spheres}
\end{figure}%

Let us consider Figure~\ref{fig-con-n-spheres} where we show a configuration of concentric spheres constructed
so that the ratio of the radii of any two adjacent spheres
is a constant, say $b/a=c$.
For $c>1$, the radii of the concentric spheres can be represented
using the sequence 
\begin{equation}
a, ca, c^2a, c^3a, \ldots,
\label{cg1outw}
\end{equation}
where the ordering is such that the radii increases to the right.
The Casimir energy of such a configuration is given by
\begin{eqnarray}
E(a,c) &=& \frac{A(c)}{a} +\frac{A(c)}{ca} +\frac{A(c)}{c^2a} + \ldots 
\nonumber \\ &=& \frac{A(c)}{a} 
\Big[1 +\frac{1}{c} +\frac{1}{c^2} + \ldots \Big] \nonumber \\
&=& \frac{A(c)}{a} \frac{c}{c-1}.
\end{eqnarray}

Observe that the Casimir energy for such a configuration of spheres
does not change sign. Again, for $c>1$, we can build another configuration
whose radii are given by the~sequence
\begin{equation}
\ldots, \frac{a}{c^3}, \frac{a}{c^2}, \frac{a}{c}, a,
\label{cg1inw}
\end{equation}
where the convention of ordering of radii is maintained.
The Casimir energy of such a configuration is given by
\begin{eqnarray}
E(a,c) &=& \frac{A(c)}{a} 
\Big[c +c^2 + c^3 + \ldots \Big] 
= -\frac{A(c)}{a} \frac{c}{c-1}.
\hspace{5mm}
\end{eqnarray}

Thus, the Casimir energy of the configuration of spheres given by
Equation\,(\ref{cg1inw}) changes sign but has the same magnitude
as the configuration in Equation\,(\ref{cg1outw}).

This counterintuitive result of the Casimir energy changing sign
has been obtained only after we replaced a divergent series with
a regularized sum, which is not conventional. It is not uncommon
for a divergent series involving a sum of positive numbers to
yield a negative regularized sum. The same behavior was seen for parallel plates.

\subsection*{Inversion}

Inversion is a transformation of a point with respect to a sphere
of radius $a$ such that the radial distance $r$ of the point from
the center of the sphere and the corresponding radial distance
$r^\prime$ of the transformed point satisfy
\begin{equation}
rr^\prime =a^2,
\end{equation}
while keeping the angular coordinates of the point unchanged.
Inversion transforms the points on a sphere into the points on
another sphere or a plane. Planes in this context are imagined
to be spheres of infinite radius.
Inversion also preserves angles and is thus closely related to 
conformal transformations. Inversion is also at the heart of the
technique of method of images in electrostatics.

How does the Casimir energy of two concentric spheres transform
under inversion about one of the spheres? The original configuration
has the energy
\begin{equation}
E(a,b) = \frac{A(c)}{a}, \qquad c=\frac{b}{a},
\end{equation}
and the Casimir energy after inversion is
\begin{equation}
E\left(\frac{a^2}{b},a\right) = \frac{A(c)}{a^2/b}.
\end{equation}

Together, this involves a set of three concentric spheres of radii
\begin{equation}
\frac{a}{c}, a, ca,
\end{equation}
whose Casimir energy is given by
\begin{equation}
E(a,c) = \frac{A(c)}{a} (c+1).
\end{equation}

Adding one more sphere with its inverse pair leads to
\begin{equation}
\frac{a}{c^2}, \frac{a}{c}, a, ca, c^2a,
\end{equation}
whose Casimir energy is given by
\begin{equation}
E(a,c) = \frac{A(c)}{a} \left[c^2+c+1 +\frac{1}{c} \right].
\end{equation}

It is easily verified that the configuration of concentric spheres
given by the sequence in Equation\,(\ref{cg1outw}), when inverted about
the innermost sphere, gives a configuration given by the sequence
in Equation\,(\ref{cg1inw}). Observe that the sphere of radius $a$ under
this inversion remains unchanged. Together, we have
\begin{equation}
\ldots, \frac{a}{c^3}, \frac{a}{c^2}, \frac{a}{c}, a,
ca, c^2a, c^3a, \ldots.
\label{invseta}
\end{equation}

It should be noted that this sequence is not the superposition
of the two sequences in Equations\,(\ref{cg1outw}) and (\ref{cg1inw}),
because an exact superposition would count the sphere of radius $a$ twice.
The Casimir energy of the configuration in Equation\,(\ref{invseta})
involves the series
\begin{equation}
\ldots +c^2 +c +1 +\frac{1}{c} +\frac{1}{c^2} +\ldots =0.
\end{equation}

It is instructive to arrive at this result using an independent method.
Let us interpret the configuration as an interaction between two bodies,

\begin{eqnarray}
\text{Object 1}&:& \quad\ldots, \frac{a}{c^3}, \frac{a}{c^2},\frac{a}{c},a, \\
\text{Object 2}&:& \quad ca, c^2a, c^3a, \ldots. \hspace{20mm}
\end{eqnarray}

The Casimir energy of two interacting bodies is constructed out of
\begin{eqnarray}
E_1 &=& -\frac{A(c)}{a} \frac{c}{c-1}, \\
E_2 &=& \frac{A(c)}{a} \frac{1}{c-1}, \\
E_\text{int} &=& \frac{A(c)}{a},
\end{eqnarray}
and is given by
\begin{equation}
E = E_1 +E_2 +E_\text{int} =0.
\end{equation}

\section{Quasi-Periodic Configuration of Plates}\label{S4}
%
An algorithm to generate a quasi-periodic configuration of plates,
using plates that satisfy Dirichlet ($D$) and Neumann ($N$) boundary conditions
on them, is the following. Start with a $D$ plate. In each iteration,
carry out the following two rules,
\begin{subequations}
\begin{eqnarray}
D(z) &\to& N(z), \\
N(z) &\to& D(z) N\left( z+ \frac{a}{2^{n-2}} \right),
\end{eqnarray}
\end{subequations}
which is read as follows: replace the Dirichlet plate at position $z$
with a Neumann plate at position $z$,
and replace a Neumann plate at position $z$ with a Dirichlet plate
at position $z$ and a Neumann plate at position $z+a/2^{n-2}$,
where $n$ represents the $n$-th iteration.
\begin{subequations}
\begin{eqnarray}
\text{Iteration $1$} &:& D, \\
\text{Iteration $2$} &:& N, \\
\text{Iteration $3$} &:& DN, \\
\text{Iteration $4$} &:& NDN, \\
\text{Iteration $5$} &:& DNNDN, \\
\text{Iteration $6$} &:& NDNDNNDN, \\
\text{Iteration $7$} &:& DNNDNNDNDNNDN, \hspace{10mm} \\
&& \hspace{16mm} \vdots  \nonumber
\end{eqnarray}
\end{subequations}

We have to also give a rule for specifying the distances between the plates.
We assume infinitely thin plates. In iteration 3, we let the distance
between the Dirichlet and Neumann plates  be $a$.

\begin{figure}[h]
\begin{center}
\includegraphics[width=7.5 cm]{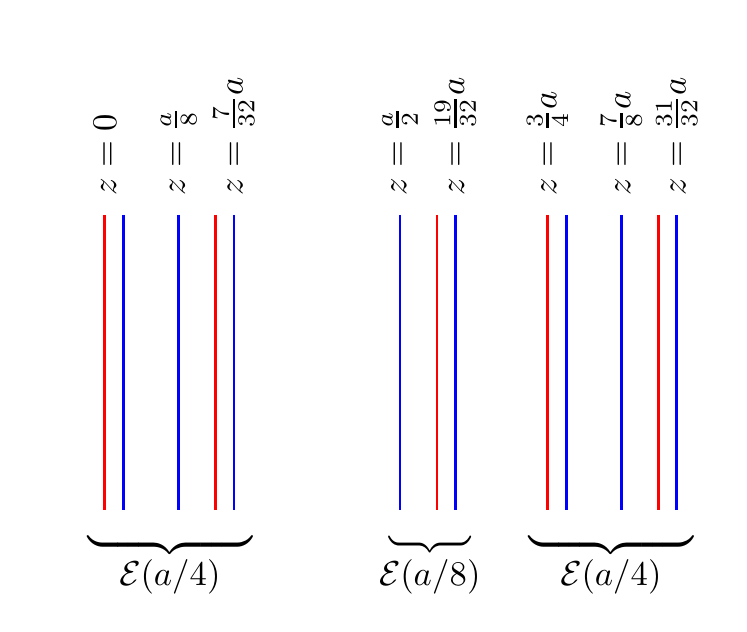}
\caption{{Quasi-periodic} configuration of plates, 7-th iteration ($n=7$). Red color represents Dirichlet plate and the blue color is Neumann plate.}
\label{fig-quasi-DN-plates}
\end{center}
\end{figure}%

A quasi-periodic configuration like the one shown in Figure~\ref{fig-quasi-DN-plates} is not periodic, even though it gives
the impression that it is periodic. We can calculate the Casimir energy
of this system using the idea of self-similarity and scaling to be
\begin{eqnarray}
{\cal E}(a) &=& 
2{\cal E}\left( \frac{a}{4} \right) +{\cal E}\left( \frac{a}{8} \right)
+{\cal E}_\text{NN}\left( \frac{a}{4} \right)
+{\cal E}_\text{ND}\left( \frac{a}{8} \right) \nonumber \\
&=& 640 {\cal E}(a) + 64 {\cal E}_\text{NN}(a) +512 {\cal E}_\text{ND}(a),
\end{eqnarray}
where ${\cal E}_\text{NN}(a)$ is the Casimir energy of two plates
satisfying Neumann boundary conditions and
where ${\cal E}_\text{ND}(a)$ is the Casimir energy of two plates
satisfying Dirichlet boundary conditions, when they are separated by distance $a$.
This implies 
\begin{equation}
{\cal E}(a) = -\frac{64}{639} {\cal E}_\text{NN}(a) 
-\frac{512}{639} {\cal E}_\text{ND}(a). 
\end{equation}

\section{Spectral Functions Revisited}\label{S5}

In the situations studied until now, a crucial factor that has made possible the application of the formalism presented above is to previously have computed the vacuum energy of the elementary building blocks that make up the self-similar structure under consideration. For the different configurations with Dirichlet plates, the Casimir energy interaction between two perfectly conducting plates was known. Likewise, for the Sierpinsky triangle, we knew the Casimir energy of the triangular cavity beforehand. These energies were, of course, only dependent on a geometric parameter that made it easier to apply the formalism. In any case, the previously calculated vacuum energy  was very well defined by a Laplacian operator in a Minkowskian space-time. We shall like to review in this section some of these concepts that,  in the next section, we will again show for a fractal space from the work of some authors (essentially working from a pure mathematical point of view).

One way to reproduce the Casimir energy interaction between two Dirichlet plates\footnote{Casimir computed this for the first time in 1948 in a completely different manner \cite{Casimir_1948}.} is by setting the differential equation satisfied by the scalar field and imposing the appropriate boundary conditions on the possible solutions, 
\begin{equation}
    \nabla^2\phi+\omega^2\phi=0,
\end{equation}
with the constraint $\phi=0$ at the boundaries. The normal modes are then given by
\begin{equation}
    \omega_n=\Bigg[k^2+\Big(\frac{n\pi}{a}\Big)^2\Bigg]^{1/2},
\end{equation}
where $n$ is a positive integer number, $k$ is the transverse momentum, and $a$ is the distance between the plates. Following Equation~(\ref{zero-point}), the zero-point energy per unit area is
\begin{equation}
    {\cal E}=\frac{1}{2}\sum_{n=1}^\infty\int\,\frac{d^2k}{(2\pi)^2}\Bigg[k^2+\Big(\frac{n\pi}{a}\Big)^2\Bigg]^{1/2}.
\end{equation}

A common way to regulate this divergent expression is using the analytic properties of the zeta-function \cite{Elizalde_book}. Changing the $1/2$ exponent by $-s/2$ and for sufficiently large $s$, the integral converges and can be calculated:
\begin{equation}
{\cal E}=\frac{1}{4a}\Big(\frac{\pi}{a}\Big)^{1-s}\frac{\Gamma[(s-2)/2]}{\Gamma(s/2)}\sum_{n=1}^\infty n^{2-s}.
\label{CasEn_zeta}
\end{equation}

The last term can be identified as a zeta function on the complex plane \cite{Elizalde_book}:
\begin{equation}
    \zeta(s)=\sum_{n=1}^\infty\,n^{-s},\qquad Re(s)>1.
\end{equation}

When we apply this to Equation~(\ref{CasEn_zeta}), the zeta function, $\zeta(s-2)$, appears, where $s$ has to be evaluated at $s=-1$. At this point the sum in Equation~(\ref{CasEn_zeta}) does not exist, but by writing it in terms of the zeta function as $\zeta(s-2)$ and then setting $s=-1$, it is possible to use analytic properties of the zeta function and obtain a finite value for $\zeta(-3)$, since this function is analytic in the whole complex plane except when it is evaluated at $1$, where it has a pole. In general, the $\zeta$ function of the Laplacian operator admits analytic continuation to the whole complex space with at most simple poles at 
\begin{equation*}
    s=\frac{d}{2}, \frac{d}{2}-1, \frac{d}{2}-2, \dots , \frac{d}{2}-\frac{d-1}{2},
\end{equation*}
where $d$ is the dimension of the space.

This, at first glance, simple way to extract the finite part of the Casimir energy hides the divergences associated with the vacuum. However, there is another spectral function called heat kernel\footnote{{It obeys the heat conduction equation, and once the initial condition is fixed, it obeys the same boundary conditions as} $\phi$.}, $K(t)$, which we can define by means of a parameter $t$ (time in the heat transfer equation) \cite{Kirsten_book},
\begin{equation}
    K(t)=\sum_n e^{-\omega_n t},
    \label{heat_kernel_expan}
\end{equation}
and relate it to the zeta function by the inverse Mellin transform,
\begin{equation}
    K(t)=\frac{1}{2\pi\,i}\int_C ds t^{-s}\Gamma(s)\zeta(s),
\end{equation}
where $C$ is an appropriate contour of integration in the vertical line of an analytic region. What is interesting is that the integration will pick up the residues due to the poles of the zeta function and Gamma function, and the resulting expression corresponds to the asymptotic behavior for small $t$ of the heat kernel,
\begin{equation}
    K(t)=\Big(\frac{1}{4\pi t}\Big)^{3/2}\Big[a_0+a_{1/2}\, t^{1/2}+a_1 \,t+a_{3/2}\,t^{3/2}+O(t^2)\Big].
\end{equation}

The coefficients in this expansion have a universal geometrical meaning, printing, in this way, the signature of the space where they live. For example, the first coefficient $a_0$ is the volume and the second one $a_{1/2}$ the surface\footnote{{In the particular case of parallel plates, these two coefficients are infinite, but this is not a problem since the background vacuum needs to be subtracted and contributes with the same terms.}}. Another spectral function, less studied but very suitable to bring out the space properties of the system, is the cylinder or Poisson kernel, which we will not consider here \cite{fulling2007}.

From Equations~(\ref{zero-point}) and (\ref{heat_kernel_expan}), we can write the total expression for the vacuum energy~as
\begin{equation}
    {\cal E}=-\frac{1}{2}\frac{d}{dt}\sum_n e^{-\omega_nt}\Big|_{t=0}
\end{equation}
where technically, the Casimir energy is defined by the finite part of this series. It is straightforward to see that the total vacuum energy is expanded in a series of divergent terms with coefficients coming from asymptotic heat kernel expansion. Therefore the same geometrical meaning can be given to the corresponding terms in the total vacuum energy. 

The same concepts can be applied to other geometries different from parallel plates, for example, a cylindrical cavity with perfectly conducting boundary conditions on the borders. Since the Laplacian is formally the same (notice that in this case, only one coordinate is infinity, and therefore there is space invariance in just the longitudinal direction), and the eigenvalues are formally looking alike,
\begin{equation}
    \omega_n=\Big[k^2+\lambda_{nm}^2\Big]^{1/2}, \label{eigenvalues}
\end{equation}
the same kind of calculation follows and one can calculate the total vacuum energy \cite{Abalo_PRD82}. In particular, for an equilateral triangle of side $a$, the eigenfunctions can be found in References~\cite{Milton_EM_waves_book_2006, Schwinger_classical_Ed_book_1998} and have a degeneracy of $6$, and $\lambda_{mn}$ in Equation~(\ref{eigenvalues}) is 
\begin{equation}
    \lambda=\frac{\pi^2}{a^2} (m^2+mn+n^2)
\end{equation}

The integers $m$ and $n$ take positive or negative values depending on the boundary condition \cite{Abalo_PRD82}. The total energy then can be read from the same reference as
\begin{equation}
    {\cal E}_\Delta= \frac{0.0237}{a^2}+\lim_{t\rightarrow 0}\Big(\frac{3\sqrt{3}}{4\pi^2}\frac{A}{t^4}-\frac{1}{8\pi}\frac{P}{t^3}+\frac{C}{48\pi t^2}\Big),
\end{equation}
where the first term is the finite Casimir energy, $A=a^2/2$ is the area of the triangle, $P=3a$ is the perimeter, and $C$ is a constant, independent of $a$ and reflects the divergences due to the corners. For an equilateral triangle $C=8$. The same behavior is supposed for the calculation in \cite{shajeshPRD96}.

This expansion seems to have a universal behavior, meaning that the coefficients of the divergences associated have an specific meaning, reminiscent of the geometry under study. This series is called the Weyl expansion \cite{Weyl_1911}, and for dimension $d$ and small t, it can be written as
\begin{equation}
   K(t)\sim\frac{1}{(4\pi t)^\frac{d}{2}}\Big[V + \alpha S \sqrt{t}+\cdots\Big]
   \label{weyl_expansion}
\end{equation}

\section{Self-Similar Manifolds}\label{S6}
%
The concept of diffusion can be interpreted as a random walk. In a regular-smooth space, the random walk generates a displacement  that is proportional to the number of steps. These can be taken in any direction. In a fractal or a self-similar system (think of the Sierpinski triangle), part of the space has been removed, and the random walk can not be taken through ``anywhere'' because the available space becomes more selective. The displacement is now proportional to the number of steps raised to a power (less than one) inversely proportional to the so-called ``walk dimension'', a dimension that is characteristic of each fractal. Diffusion on a fractal space is then slower than in a regular space, ``the ant in the labyrinth'' \cite{Gennes_1976}.

Likewise, the Laplace operator $\triangle=\nabla^2$, like the one in the previous section, can be defined in a smooth manifold provided with a Riemannian metric \cite{Jost_book, Boothby_book, Molchanov_1975}. This operator is self-adjoint and 
 is used to define spectral equations like the diffusion of heat.

From several cited authors, we know that in a fractal, it is not possible to define the Laplace operator in the common fashion as we are used to do in regular space-time, because the derivative is not naturally defined in a fractal space. As a consequence, it is not possible to write differential operators directly, and other ways of defining the Laplacian have been developed; see \cite{Derfel_JPhysA45_2012} and references therein.

Therefore, spectral functions derived from differential equations defined on a fractal do not have the same properties as the ones defined on a smooth manifold, which has been subject of study among mathematicians. The Sierpinski triangle is one of the fractals studied in more detail. Using spectral decimation techniques \cite{Fukushima_1992, Shima_1991}, the eigenvalues of a self-similar Sierpinski triangle in $d$ dimensions has been calculated in Reference~\cite{Shima_1991} (for example). They find the set of eigenvalues that for a 2D Sierpinski triangle are 
\begin{equation}
    \lbrace-2, -3, -5\rbrace,
\end{equation}
with degeneracies,
\begin{eqnarray}
d_m(-2)=\left\lbrace
\begin{array}{c} 
1 \qquad\quad m=1, \\ 
0 \quad \text{otherwise.} 
\end{array}
\right.\\
d_m(-3)=\frac{1}{2}(3)^{m-1}-\frac{3}{2}, \qquad m\geq 2\\
d_m(-5)=\frac{1}{2}(3)^{m-1}+\frac{3}{2}, \qquad m\geq 1.
\end{eqnarray}

Notice that the degeneracies grow exponentially, while in a regular manifold, they are, at most, polynomials. In general, the same thing happens for the eigenvalues.
The $\zeta$ function for the Sierpinski triangle is calculated in Reference~\cite{derfel_2008, Teplyaev_2007, Kigami_1989} following a spectral decimation technique. The result can be expressed as
\begin{eqnarray}
    \zeta_\triangle(s)&=&5^{-s}\zeta_{\phi,-2}+\Bigg[\frac{3}{2(5^s-3)}-\frac{3}{2(5^s-1)}\Bigg]\,5^{-s}\zeta_{\phi,-3}(s)+\nonumber\\
    &+&\Bigg[\frac{1}{2(5^s-3)}+\frac{3}{2(5^s-1)}\Bigg]\,\zeta_{\phi,-5}(s)
    \label{zeta-triangle}
\end{eqnarray}
where the $\zeta_{\phi,\lambda}$ are analytically continued to Real $(s)<0$ and have only simple poles \cite{Derfel_JPhysA45_2012}.
Moreover, rather than having real poles as in the regular space, they are complex poles,
\begin{equation}
    s=\frac{d_s}{2}+\frac{2\pi\,i\,m}{log 5}, \qquad m \in \mathcal{Z},
\end{equation}
where $d_s$ is the spectral dimension.  References~\cite{Teplyaev_2007, Dunne_JPhysA45_2012} sketch the distributions of the poles in the complex plane where the complex poles appear as a dot-tower in the vertical line ``whose real part defines the spectral dimension $d_s/2$''.

The heat kernel for small $t$ in a fractal system behaves very differently from the asymptotic behavior found in Equation~(\ref{weyl_expansion}). In general, for a self-similar structure, the heat kernel is bounded. An upper bond and a lower bond is calculated in \cite{Kumagai_1993}. In general, it is possible to write
\begin{equation}
    C_1t^{-d_s/2} \leq K(t) \leq C_2t^{-d_s/2},
\end{equation}
for some constants $C_1$ and $C_2$. The expansion of the heat kernel for small $t$ is $t^{-d_s/2}$ to leading order, but rather than behaving like in a Riemannian space, the presence of complex poles give rise to periodic oscillations in the next terms. The exponents in the expansion depend on  the spectral dimension $d_s$ and the walk dimension $d_w$, which  are related with the Hausdorff dimension $d_h$ by means of $d_s\,d_w = 2\,d_h$. In the usual space of a smooth manifold $d_w=2$ and the spectral dimension coincides with the Hausdorff dimension. Physical interpretation and consequences of the appearance of the oscillatory behavior of the heat kernel can be found in References~\cite{Akkermans_2009,Akkermans_PRL_105_2010}. 

\subsection*{Example \textendash Casimir Energy of a Sierpinski Triangle}

We report here the result from \cite{Derfel_JPhysA45_2012}, where they calculate the Casimir energy of a fractal. In particular, they compute it for a Sierpinski triangle with Dirichlet and Neumann boundary conditions using spectral decimation. Once they have the $\zeta_\triangle$ function for the Sierpinski triangle shown above in Equation~(\ref{zeta-triangle}), they can proceed to calculate the energy in a standard manner.

They follow the standard procedure in \cite{Kirsten_book} and start by calculating the zeta function of the differential operator
\begin{equation}
    P=-\frac{\partial^2}{\partial \tau^2}+\nabla^2,\qquad  \text{on}\quad (\mathbb{R}/\beta^{-1}{\mathbb{Z}})\times G,
\end{equation}
where \emph{G} is the self-similar Sierpinski triangle and $\beta=1/kT$. Then, the total Casimir energy is defined by
\begin{equation}
    E=-\frac{1}{2}\frac{\partial}{\partial\beta}\zeta'_{P/\mu^2}(0),
    \label{CasEn_del_beta}
\end{equation}
where the re-scaled operator $P/\mu^2$ is considered.

Imposing Dirichlet boundary conditions on the fractal space and periodic boundary conditions on the $\tau$ variable, we get an extra term for the eigenvalues,
\begin{equation}
    \Big(\frac{2\pi\,n}{\beta}\Big)^2,
\end{equation}
coming from the periodicity on time, so that the $\zeta_P$ function of the $P$ operator, in terms of the Mellin transform, can be writen as,
\begin{equation}
    \zeta_P(s)=\frac{1}{\Gamma(s)}\sum_{n=-\infty}^{\infty}\int_0^\infty\frac{dt}{t}t^{s-1} e^{-(\frac{2\pi\,n}{\beta})^2t}\,K(t).
\end{equation}

Using Poisson resummation,
\begin{equation}
    \sum_{n=-\infty}^{\infty}e^{-(\frac{2\pi\,n}{\beta})^2t}= \frac{\beta}{2\sqrt{\pi\,t}}\sum_{n=-\infty}^{\infty}e^{-\frac{\beta^2\,n^2}{4t}}
\end{equation}
and isolating the term $n=0$, we find,
\begin{equation}
    \zeta_P(s)=\frac{\beta}{2\sqrt{\pi}\Gamma(s)}\Gamma\Big(s-\frac{1}{2}\Big)\zeta_\triangle(s-\frac{1}{2}\Big)+\frac{\beta}{\sqrt{\pi}\Gamma(s)}\int_0^\infty dt\, t^{s-\frac{3}{2}}\sum_{n=1}^{\infty}e^{-\frac{\beta^2\,n^2}{4t}}.
\end{equation}

For the re-scaled operator $P/\mu^2$, the above expression gets multiplied by $\mu^{2s}$ so that the derivative that shows in Equation~(\ref{CasEn_del_beta}) can be written as,
\begin{equation}
    \zeta'_{P/\mu^2}(0)=\zeta'_P(0)+\zeta_P(0)\,\ln\mu^2.
\end{equation}
    
    The second term is zero and the first term is
\begin{equation}
       \zeta'_P(0)=-\beta\zeta_\triangle\Big(-\frac{1}{2}\Big)-2\,\sum_{j=1}^\infty\ln\left(1-e^{-\beta\sqrt{\lambda_j}}\right).
\end{equation}

Performing the derivative of that expression evaluated at zero and making $\beta\rightarrow\infty$ to be at zero temperature, the Casimir energy is found to be,
\begin{equation}
   E= \frac{1}{2}\zeta_\triangle\Big(-\frac{1}{2}\Big)
\end{equation}

This has to be evaluated using Equation~(\ref{zeta-triangle}) in terms of the $\zeta$ functions $\zeta_{\phi,\lambda}$ that are given in Reference~\cite{Derfel_JPhysA45_2012}. There, they do the numerical calculation and find
a numerical value for the energy.
Similarly, they give an estimation for the Casimir energy of the Sierpinski triangle with Neumann boundary conditions
.

\section{Discussion}\label{S7}
%
We have studied the vacuum energy in self-similar systems distinguishing two ways of constructing such systems. The repetitive nature of a self-similar system allows one to consider a recursive element and arrange them in a similar-looking manner, such that the outcome reveals a pattern that repeats at different scales. In this sense, a self-similar system exhibits scale invariance. Under these terms, we have considered several self-similar structures in Sections \ref{S2}--\ref{S4}, which are treated as a many-body problem using the multiple scattering approach. We have studied the Dirichlet boundary conditions for the simplicity in the calculations, but we could follow the same approach for Neumann boundary conditions. The propagating fields considered in Section \ref{S2} propagate in a regular manifold in a Minkowski space-time. There, a pile of generating objects with different scale rearrange in the right way to look like a Sierpinski carpet. The objects have perfectly conductor boundaries that divide the regular space-time in sectors which overall looked like Sierpinski object.

In Section \ref{S6}, however, the space itself is self-similar. It is on this self-similar manifold where the fields propagate carrying with them the anomalous signature that characterizes a fractal in general. This anomalous propagation is reflected on the eigenfunctions and eigenmodes of a Laplacian that, defined on a fractal space, can not have the same properties as the Laplacian defined on a Riemannian space. As a consequence, the dimension of fractal space is not unique but characterizes a property of such space. The Hausdorff dimension, $d_h$, (also known as fractal dimension) makes reference to the geometric distribution, or scaling properties of the space, while the spectral dimension $d_s$ characterizes the scaling properties of the eigenvalues of the Laplacian that has been defined on the fractal space. Finally, the concept of random walk introduces another dimension $d_w$, which is no longer independent of the other two, $d_w\,d_s = 2d_h$. In a smooth regular space, $d_w=2$ and the spectral dimension and Hausdorff dimension coincide.

The discussion above has immediate consequences in the physics that each of the phenomena represent. The quantum vacuum in a Riemaniann manifold has been broadly studied and its manifestations under external conditions widely investigated. Spectral functions have provided an understanding of the inner connection between the vacuum fluctuations and the space where they manifested. In this sense,  the asymptotic behavior of the heat kernel expansion for small time and the properties of the zeta function (which derivative is proportional to the vacuum energy) have been investigated by the community. In a regular manifold, the zeta function admits continuation to the whole complex plane by studying the asymptotic behavior of the heat kernel expansion. The residues of the poles of the zeta function have geometric meaning. Some authors have tried to extend this to a fractal space. Several difficulties arise, such as how to determine the Laplacian in these spaces. The Weyl expansion, which comes as a consequence of the heat kernel expansion of $t$ with powers dependent on the dimension of the space,  has a universal behavior. In the case of fractal space, Berry conjectured that the same behavior of the heat expansion would be found where the dimension of the space would now be the Hausdorff dimension. This has been proven not to be the case \cite{Brossard_1986}. However, since the fractal dimension is not the only dimension that can be defined on a fractal, other possibilities, considering spectral dimension, have also been investigated, but no firm conclusion is established yet \cite{Lapidus_1991, Lapidus_1993, Lapidus_1996}.

It has been pointed out that the heat kernel in a fractal space has a different behavior than in the regular space. The existence of complex poles of the zeta function generated periodic behavior in high order corrections. The first order goes like the negative power of the spectral dimension rather than of the fractal dimension. Moreover, the heat kernel does not converge but  is bounded from above and below.

We have illustrated how the two ways of investigating the Casimir effect for  the Sierpinski triangle with Dirichlet boundary conditions are different in nature since they are considered in different spaces. In our approach, we are able to calculate the energy of self-similar configurations built up from objects (building blocks) whose Casimir energy can be computed and a geometric relation can be established. We have calculated the vacuum energy of structures that cover the whole space and systems that mimic quasi-periodic configurations. In future work, we plan to further develop the system that in \cite{shajeshPRD96} we call ``Inverse Sierpinski triangle''.
\vspace{6pt}

\section*{ Acknowledgment}
I.C-P. is grateful for support of the GOBIERNO DE ARAG\'ON grant number E21\_20R and FONDOS FEDER---AGENCIA ESTATAL DE INVESTIGACI\'ON grant number PGC2018-095328-B-I00.

\bibliographystyle{unsrt}

\end{document}